# Electrical Control of Structural and Physical Properties via Strong Spin-Orbit Interactions in $Sr_2IrO_4$

G. Cao[1*], J. Terzic[1], H. D. Zhao[1], H. Zheng[1], L. E. De Long[2], and Peter S. Riseborough[3]

[1] Department of Physics, University of Colorado at Boulder, Boulder, CO 80309

[2] Department of Physics and Astronomy, University of Kentucky, Lexington, KY 40506

[3] Department of Physics, Temple University, Philadelphia, PA 19122

Electrical control of structural and physical properties is a long-sought, but elusive goal of contemporary science and technology. We demonstrate that a combination of strong spin-orbit interactions (SOI) and a canted antiferromagnetic (AFM) Mott state is sufficient to attain that goal. The AFM insulator $Sr_2IrO_4$ provides a model system in which strong SOI lock canted Ir magnetic moments to $IrO_6$-octahedra, causing them to rigidly rotate together. A novel coupling between an applied electrical current and the canting angle reduces the Néel temperature and drives a large, non-linear lattice expansion that closely tracks the magnetization, increases the electron mobility, and precipitates a unique resistive switching effect. Our observations open new avenues for understanding fundamental physics driven by strong SOI in condensed matter, and provide a new paradigm for functional materials and devices.

[*] *Corresponding author: gang.cao@colorado.edu*

It is now widely recognized that a unique competition between spin-orbit interactions (SOI) and Coulomb correlations, U, in *4d-* and *5d-*elements and their compounds drives unusual physical behaviors that markedly differ from those of their *3d* counterparts (1,2). The *5d*-iridates display particularly strong and surprising influences of SOI on their physical properties. Early studies (3,4,5,6,7,8) indicated iridates exhibit a preference for magnetic, insulating ground states, a trend now recognized as a consequence of a combined effect of U and strong SOI. An important example of this effect is the $J_{eff}$ = 1/2 Mott state identified in $Sr_2IrO_4$ (9), whose defining characteristic is the strong locking of the lattice and Ir magnetic moments (1,2,10).

We have conducted a new study of $Sr_2IrO_4$ that centers on unconventional, single-crystal x-ray diffraction measurements with ***simultaneous application of electrical current to diffracted samples,*** as well as the I-V characteristics, electrical resistivity and magnetization as functions of temperature, electrical current and magnetic field. Our central finding is that application of electrical current causes the ***a***-axis lattice parameter to expand by an astonishing 1% that, in turn, precipitates profound changes in physical properties. The current-controlled lattice expansion closely tracks the long-range magnetic order, causing a considerable decrease in both the Néel temperature $T_N$ and magnetization, due to the strong SOI that rigidly locks the Ir moments to the lattice. The current-dependence of the *a*-axis expansion is highly non-linear, which induces the novel I-V characteristics of $Sr_2IrO_4$.

Simultaneous control of structural and physical properties via electrical current is a rare, but extremely desirable contemporary goal because of its great technological potential. Our discovery of such behavior in $Sr_2IrO_4$ opens new avenues for understanding the fundamental consequences of strong SOI in crystalline solids, and provides a new paradigm for development of functional materials and devices.

An explanation of experimental techniqueps and additional discussion are presented in the Supplemental Material (SM) (11).

Sr$_2$IrO$_4$ is the archetype, SOI-driven insulator (9) with $T_N$ = 240 K (3,4,5,6), and an electronic energy gap $\Delta \leq 0.62$ eV (9,12,13,14,15). It crystallizes in a tetragonal structure with space-group *I4$_1$/acd* (No. 142) with $a = b = 5.4846$ Å and $c = 25.804$ Å at temperature T = 13 K (3,4,5). Recent studies suggest a further reduced space group *I4$_1$/a* (No. 88) for Sr$_2$IrO$_4$ (16,17). Two signature characteristics of Sr$_2$IrO$_4$ are essential for understanding the results of this study: **(1)** Rotation of the IrO$_6$-octahedra about the *c*-axis by approximately 12°, which corresponds to a distorted in-plane Ir1-O2-Ir1 bond angle θ, has a critical effect on the ground state (2,18,19,20,21,22,23,24,25). **(2)** The magnetic structure is composed of ordered moments (0.208(3) μ$_B$/Ir) canted ***within the basal plane*** (16). The 13(1)° canting of the moments away from the *a*-axis closely tracks the staggered rotation of the IrO$_6$ octahedra (26), which sharply contrasts the behavior of *3d* oxides (27).

A representative diffraction pattern taken with the basal plane of Sr$_2$IrO$_4$ aligned nearly perpendicular to the incident x-ray beam is shown in **Fig. 1a**. The Bragg peaks for Miller indices (220) or (0016) are representative for the discussion that follows. The position and intensity of the (0016) peak for temperature T = 200 K are shown in **Fig. 1b** and **Fig. 1c**, respectively, and undergo remarkable changes upon the application of basal-plane electrical currents I up to 105 mA. The (0016) peak shifts up and to the right with a threefold reduction in intensity that is sensitive to the atomic positions within a unit cell (**Fig. 1c**). Other Bragg peaks exhibit similar shifts with I, and either enhanced or reduced intensities, which reflects changing interference generated by shifts in atom positions. The current-induced lattice changes are also accompanied by a subtle but visible color and size change of the sample, as seen with the aid of a polarizing microscope (11).

The current-controlled changes in the *a-* and *c-*axis lattice parameters were quantitatively characterized by x-ray diffraction for I applied along either the basal plane or the *c-*axis. The lattice responds more strongly to current in the basal plane than along the *c-*axis, which suggests that the orientation of the Ir moments is important, and that Joule heating is not affecting the data (11).

We now focus on normalized changes in the *a-* and *c-*axis lattice parameters $\Delta a/a$ and $\Delta c/c$ with basal-plane I, where $\Delta a/a \equiv [a(I)-a(0)]/a(0)$, and 0 mA ≤ I ≤ 105 mA; $\Delta c/c$ is similarly defined. **Figure 2a** shows $\Delta a/a$ peaks at nearly 1% near $T_N$, then decreases to 0.2 % at 300 K, whereas $\Delta c/c$ < 0.1%. The clear difference between $\Delta a/a$ and $\Delta c/c$ once again does not support a Joule heating effect (11), and further confirms an important role for the in-plane Ir moments. A more striking observation is that the temperature dependence of $\Delta a/a$ closely tracks that of the *a-*axis magnetization, $M_a$, (**Fig. 2a**); this is direct evidence that ***the current-controlled expansion of the a-axis involves interlocking of cooperative magnetic order and the lattice.*** The reduced magnetic canting must be accompanied by a simultaneous increase of θ, which is a critical parameter for determining the ground state (24, 25).

We expect the current-controlled lattice expansion to be strongly associated with long-range AFM order. We therefore undertook a parallel study of $Sr_2Ir_{0.97}Tb_{0.03}O_4$, since a 3% replacement of $Ir^{4+}$ by $Tb^{4+}$ suppresses $T_N$ to zero, but conveniently retains the insulating state and the original crystal structure (28). We found the absolute values of $\Delta a/a$ and $\Delta c/c$ for $Sr_2Ir_{0.97}Tb_{0.03}O_4$ for I = 105 mA are small (< 0.2%) and weakly temperature-dependent in the absence of AFM order (**Fig. 2b**). A comparison of **Figs. 2a** and **2b** clearly points to a critical role played by long-range AFM in the current-controlled lattice expansion, and essentially eliminates the possibility of a Joule heating effect (11).

We also examined the conventional thermal expansion of $Sr_2IrO_4$ ***measured without application of I***. The temperature dependences of the ***a-*** and ***c-***axis lattice parameters and their corresponding changes $\delta a/a$ and $\delta c/c$ due to pure thermal expansion ($\delta a/a \equiv [a(T)-a(90K)]/a(90K)$ and $\delta c/c$ is similarly defined) demonstrate that the ***a-***axis expands linearly and only slightly (~ 0.1%) from 90 K to 300 K (**Figs. 2c** and **2d**). The corresponding coefficient of linear thermal expansion $\alpha \equiv 1/a \, (da/dT)$ is approximately $5.0 \times 10^{-6}$ K$^{-1}$, which is small and comparable to those of many materials (29). The small thermal expansion of $Sr_2IrO_4$ is also consistent with its high melting point (> 1900 °C), which reflects bond energies on the order of electron volts. The sharp contrast between the conventional thermal expansion $\delta a/a$ (0.1%) and the novel current-controlled $\Delta a/a$ (~1%) highlights the extraordinary coupling between current and the AFM state.

We also observe significant changes in the ***a-***axis magnetic susceptibility $\chi_a(T)$ and the ***a-***axis magnetization $M_a$ when current is applied (**Fig. 3**). $T_N$ is drastically decreased by 40 K for I = 80 mA (**Figs. 3a** and **3b**), and the value of $M_a$ is reduced by 16% (**Fig.3c** and **3d**). Magnetic canting is ascribed to a Dzyaloshinsky-Moriya interaction (24, 30) that is closely associated with θ; the canting decreases with increasing θ and vanishes when θ = 180° (24). This is consistent with the reduced $M_a$ that signals enhanced itinerancy due to increased θ.

Another prominent consequence of the current-controlled lattice expansion is non-Ohmic behavior that features a negative differential resistance (NDR). NDR (31,32,33,34) is a nonlinear phenomenon with a negative ratio of voltage shift in response to a current change, $\Delta V \Delta I < 0$, contrary to Ohm's law, which describes the traditional positive, linear relationship, $\Delta V \Delta I > 0$. The NDR phenomenon is in general attributed to either an "electrothermal" effect or a "transferred carrier" effect (31,32). The more common form of NDR is manifest in "N"-shaped I-V characteristics (31,32,33,34). Alternatively, an "S"-shaped NDR has been observed in various memory devices (35,36) and a few bulk materials such as $VO_2$, $CuIr_2S_{4-x}Se_x$, $Ca_3Ru_2O_7$ and 1T-$TaS_2$ (36,37,38,39,40). These bulk materials are characterized by a first-order metal-insulator transition (MIT) and, except for $Ca_3Ru_2O_7$, are without an AFM state. The "S"-shaped NDR in these materials is closely associated with the first-order MIT, and attributed to drastic differences in crystal and electronic structures below and above the MIT (41). In contrast, $Sr_2IrO_4$ features strong AFM order and a Mott insulating state that persists up to at least 600 K *without a MIT* (1,2,18, 20, 21), indicating a different mechanism that drives the NDR in $Sr_2IrO_4$.

An "S"-shaped NDR was observed in an earlier study of $Sr_2IrO_4$ (6), but the underlying mechanism remained unclear up to now. The I-V curves for I applied along either the *a*- or *c*-axis at a few temperatures are presented in **Figs. 4a** and **4b**, along with the strong anisotropy of the response in **Fig. 4c**. A linear I-V response during an initial current ramp is followed by a sharp threshold voltage $V_{th}$ that marks a switching point where V abruptly drops with increasing I, thus signaling a NDR. Another broad turning point emerges with further current increase, and is more distinct in the *c*-axis I-V curves for T < 100 K.

A plot of $V_{th}$ as a function of temperature displays a pronounced slope change near 100 K, where an anomaly in $M_a$ occurs (**Fig. 4d**). Previous studies (18,21) have shown that $M_a$ undergoes two additional anomalies at $T_M \approx 100$ K and 25 K (**Fig. 4d**) due to moment reorientations, which is corroborated by a muon-spin rotation and relaxation study (19). We note that the increased scatter in the *a*- and *c*-axis parameters between 100 K and 150 K in **Figs. 2c** and **2d** are most likely due to the reorientation of the Ir moments, and are at the root of the unusual magnetoresistivity (21) and magnetoelectric behavior (18). This magnetic reorientation separates the different regimes of I-V behavior below and above $T_M \approx 100$ K, and suggests that a close relation exists between the magnetic state and the I-V characteristics (**Fig. 4d**).

We propose the NDR behavior exhibited by $Sr_2IrO_4$ reflects a novel mechanism that fundamentally differs from that operating in other materials. Our proposal is based on constructing a picture that self-consistently explains the complex NDR behavior, and the current-controlled expansion and magnetization data.

We begin by examining the *a*-axis resistivity $\rho_a$, which drops by nearly three orders of magnitude at low temperatures (**Fig. 4e**), and the corresponding activation energy gap (estimated from data covering the range 100 – 270 K), which decreases from 81 meV to 32 meV, as I increases from 0.1 mA to 20 mA. There is a clear drop of $\rho_a$ with decreasing temperature after peaking around 11 K (**Fig. 4e Inset**), indicating an incipient metallic state. The representative differential resistance, dV/dI, (11) at 100 K reveals two anomalies near 10 mA and 45 mA, marked as $I_{C1}$ and $I_{C2}$, respectively (**Fig. 4f**). Corresponding I-V curves at 100 K feature a sharp switching point ($V_{th}$) at $I_{C1}$ and a broader turning point near $I_{C2}$ (**Fig. 5a**).

***It is crucial to note that the current-controlled a-axis expansion $\Delta a/a$ closely tracks the I-V curves with non-linear changes at $I_{C1}$ and $I_{C2}$, respectively*** (upper horizontal axis in **Fig. 5a**). The slope changes in $\Delta a/a$ signal successively more rapid expansions of the *a*-axis at $I_{C1}$ and $I_{C2}$, each accompanied by a more significant increase in θ, which, in turn, enhances electron hopping (**Fig. 5b**) (more discussion below). As the current further increases above $I_{C2}$ = 45 mA, $\Delta a/a$ appears to saturate. This explains why a magnetic field H reduces V considerably only between $I_{C1}$ and $I_{C2}$, but shows no visible effect above $I_{C2}$ (see green curve in **Fig. 5a**), because H can only increase θ via realigning the Ir moments below $I_{C2}$: Above $I_{C2}$, the saturation of $\Delta a/a$ corresponds to θ approaching $180^o$, which precludes further increases, and magnetic field can therefore no longer affect the I-V curves. ***The close association between $\Delta a/a$, moment canting, and the I-V curves reveals how current-controlled basal-plane expansion drives the nonliner I-V characteristics.***

Fundamentally, the formation of the Mott insulating state with canted $IrO_6$-octahedra and canted (locked) moments is caused by a cooperative transition in which the electronic structure gaps, thereby lowering its energy relative to the paramagnetic metallic state. The gapping mechanism involves electronic correlations that involve both spin-orbit coupling and scattering through the magnetic reciprocal lattice vector. The electronic correlation is expected to manifest itself in the unoccupied states (electron-carrier) and in the occupied (hole-carrier) states. The momentum shift associated with a finite current is usually negligible in uncorrelated systems, but in correlated systems close to quantum critical points, theory shows that relatively small changes in the low-energy electronic structure can cause large (non-linear) changes in the ordered structure. In short, a slight modification of the electronic structure induced by current may result in strong modifications of the electronic correlations. ***The NDR data is interpreted in terms of a reduction in the gapping, as suggested in Fig.4e, and the concomitant decrease in the carrier effective mass induced by current.***

We have shown that a combination of strong SOI and canted AFM order can lead to a highly desirable paradigm for simultaneous electrical control of the crystal structure and physical properties of $Sr_2IrO_4$: **(1)** Strong SOI lock canted Ir moments to the $IrO_6$-octahedra, which rigidly rotate together (**Fig. 2**). **(2)** Strong SOI dictate the low-energy Hamiltonian and create small gaps in the electronic structure, which ultimately affect electron mobility (i.e., an increase in θ favors electron hopping; see **Figs. 4** and **5**). **(3)** Applied current effectively drives a lattice expansion by increasing θ and reducing small gaps in the electronic structure (**Fig. 4e**), which also reduces $T_N$ and the Ir moments (**Fig. 3**).

Our on-going research on the physical consequences of strong SOI in materials suggests that similar behavior may be widespread in other iridates. This work offers a new paradigm for studies of the physics driven by the SOI and may help unlock a world of possibilities for functional materials and devices.

**Acknowledgments**

G.C. is indebted to Drs. Xiangang Wan, Daniel Khomskii, Feng Ye and Natalie Perkins and S.-W. Cheong for stimulating discussions. This work was supported by the National Science Foundation via grants DMR-1712101 (U. Colorado) and DMR-1506979 (LED), and by Department of Energy, Office of Basic Energy Science, Materials Science through the award DEFG02-84ER45872 (PSR).9

**Captions**

**Fig. 1.** Single-crystal x-ray diffraction of $Sr_2IrO_4$ with current I applied within the basal plane. **(a)** Representative x-ray diffraction pattern of a single crystal. The circled Bragg peak is (0,0,16). **Inset**: Sample mounting showing electrical leads and cryogenic gas feed (11). Current-controlled changes in **(b)** the location and **(c)** the intensity (counts) of the (0016) peak for I = 0 and I = 105 mA.

**Fig. 2. (a)** Current-controlled shifts $\Delta a/a$ and $\Delta c/c$ for $Sr_2IrO_4$. Note that $\Delta a/a$ closely tracks $M_a$ (right scale). **(b)** For comparison to **(a)**: $\Delta a/a$ and $\Delta c/c$ for $Sr_2Ir_{0.97}Tb_{0.03}O_4$. Note that the scales for $\Delta a/a$, $\Delta c/c$ and $M_a$ are the same as those in **(a)** to facilitate comparisons. **(c)** Temperature dependence of *a*- and *c*-axis thermal expansion of $Sr_2IrO_4$ for I = 0; **(d)** Temperature-induced shifts $\delta a/a$ and $\delta c/c$ corresponding to **(c)**.

**Fig. 3.** Temperature dependence of the **(a)** *a*-axis magnetic susceptibility $\chi_a(T)$ at a few representative currents, and **(b)** $d\chi_a(T)/dT$, which clarifies the decrease in $T_N$ with I. **(c)** $M_a(H)$ at 100 K for a few representative currents. **(d)** Current dependence of $T_N$ and $M_a$. Diagrams illustrate the current-controlled lattice expansion, angle θ (red) and Ir moments (black arrows) with I.

**Fig. 4.** I-V curves for $Sr_2IrO_4$ for: **(a)** I applied along the *a*-axis, **(b)** along the *c*-axis, **(c)** both the *a*- and *c*-axis at T = 100 K. Arrows show the evolution of the current sweeps in **(a)** to **(c)**. Temperature dependence of **(d)** the threshold voltage $V_{th}$ and $M_a(T)$ (right scale), and **(e)** the *a*-axis resistivity $\rho_a$. **Inset:** Expanded $\rho_a$ for I = 20 mA. **(f)** Representative data for dV/dI as a function of DC current at T = 100 K. Note two slope changes marked at I = $I_{C1}$ and $I_{C2}$. Arrows show the sequence of applied I, and the dashed line is a guide to the eye.

**Fig. 5. (a)** I-V curves (red, blue, green) for the *a*- and *c*-axis lattice parameters at T = 100 K and $\mu_oH = 0$ and 5 T along the *c*-axis. Light blue data (upper horizontal axis) show $\Delta a/a$ at T = 100 K.

Dashed lines are guides to the eye. Note slope changes of $\Delta a/a$ occur at the two turning points at $I_{C1}$ and $I_{C2}$, respectively. **(b)** Diagrams (not to scale) illustrate the increasing lattice expansion, decreasing θ (red) and Ir moment canting (black arrows) with increasing I. Schematic increase of electron mobility due to decreased gapping of the electronic structure with I.

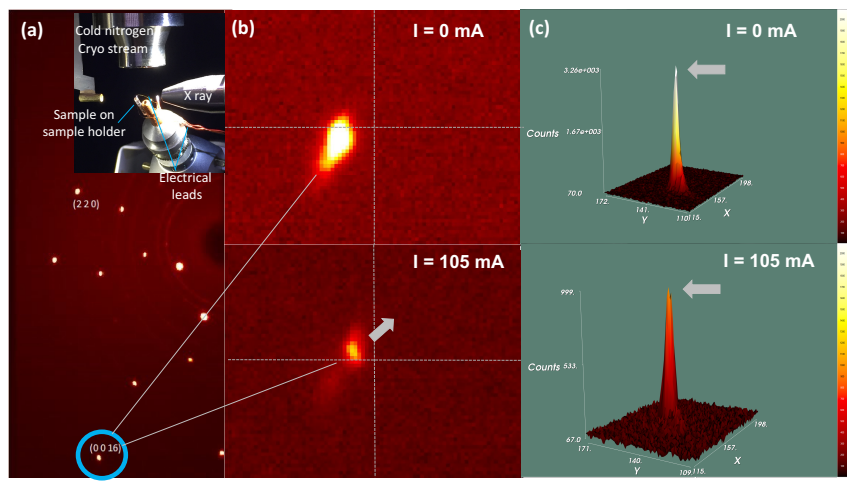

Fig. 1

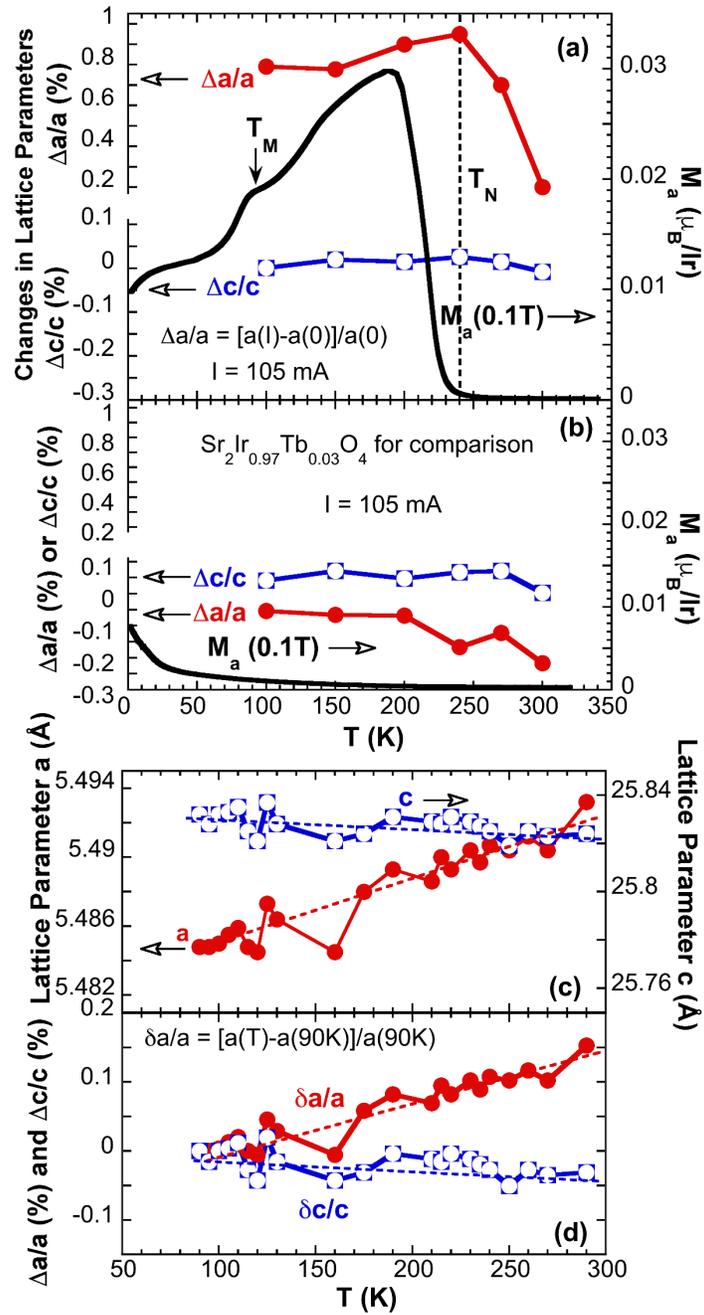

Fig. 2

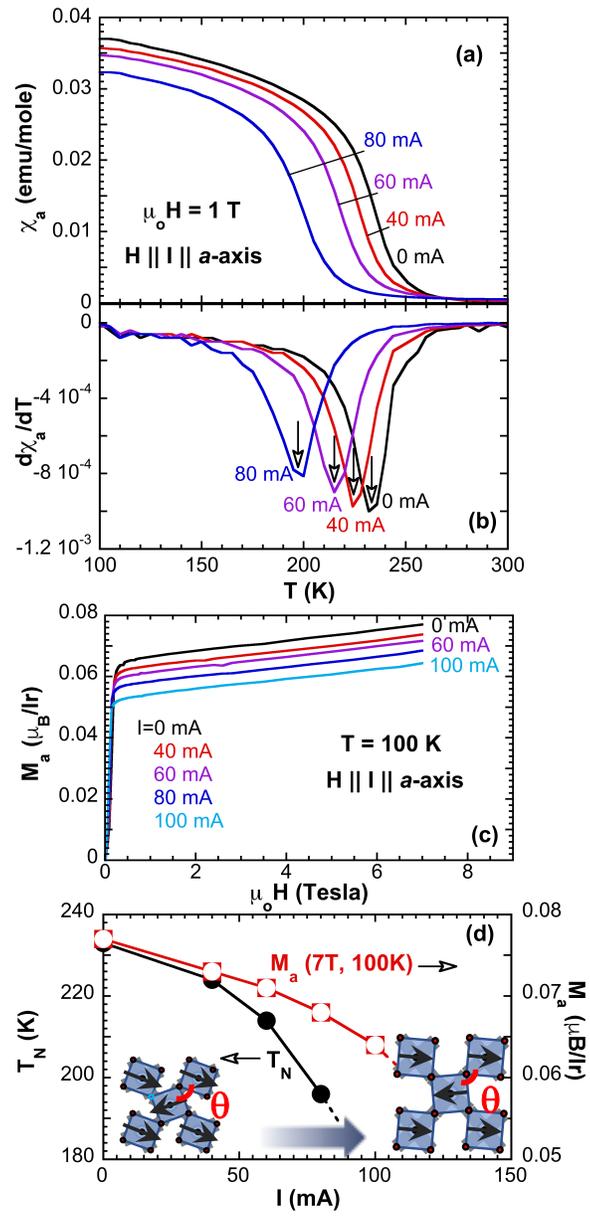

Fig. 3

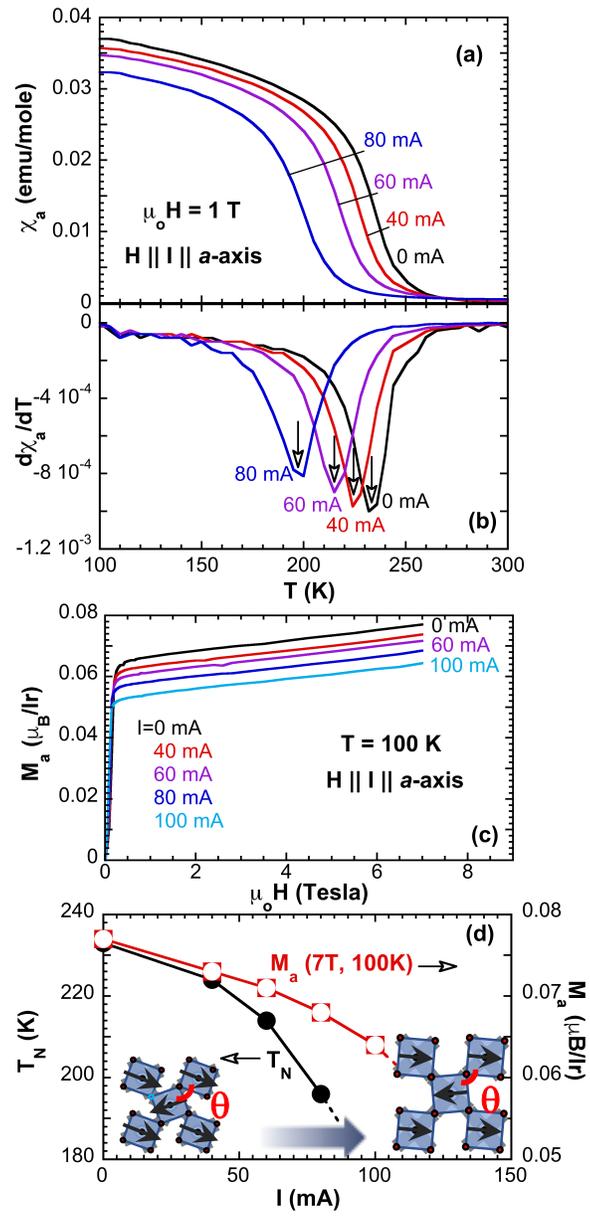

Fig. 3

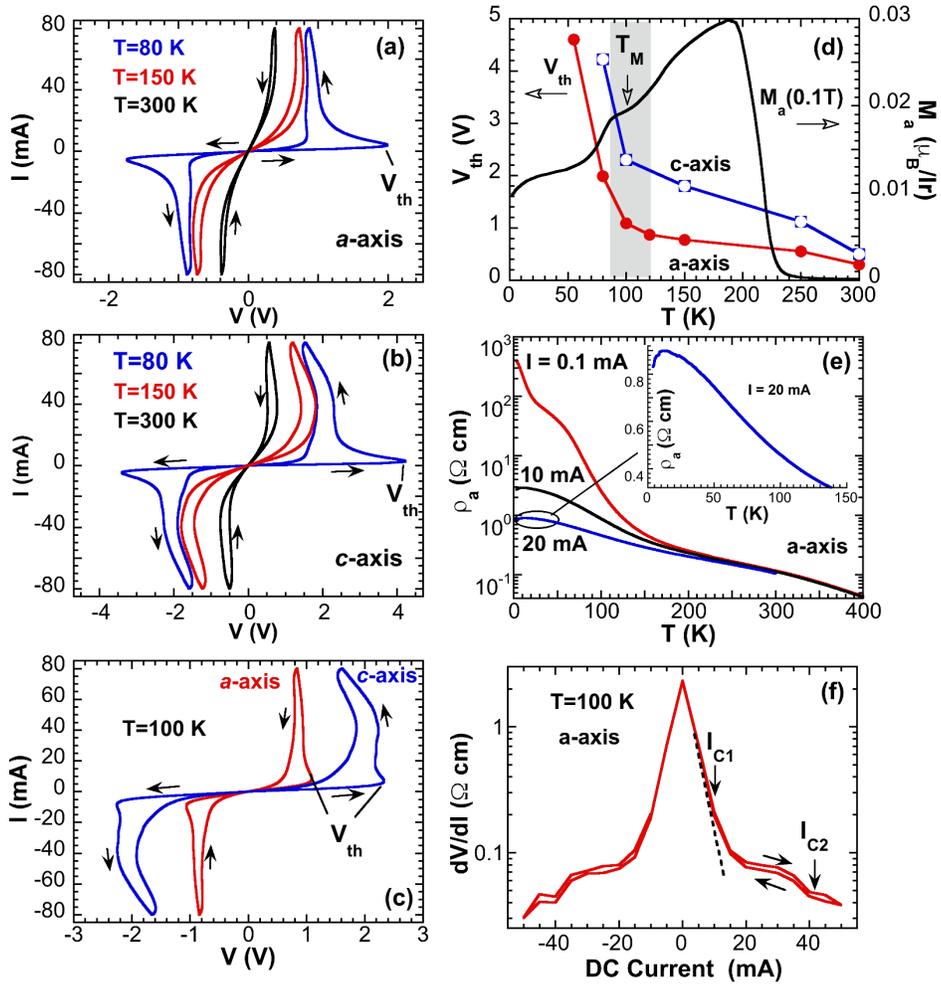

Fig.4

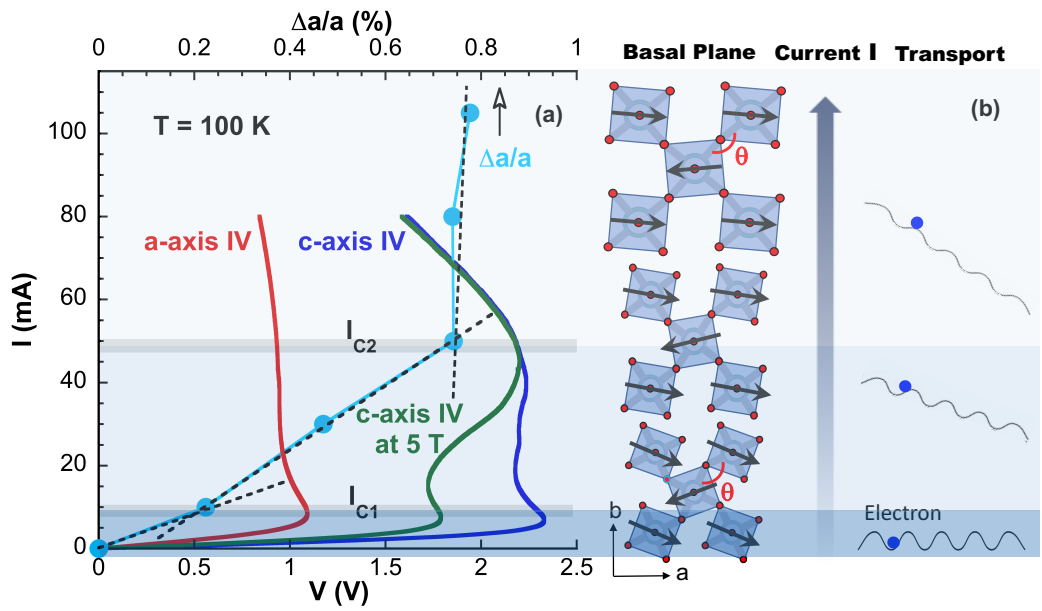

Fig.5